\documentclass[11pt]{article}

\usepackage{amsmath,amssymb,amsthm,graphicx,latexsym}
\usepackage{showkeys}
\newcommand{\ed}{\end{document}}
\newcommand{\beq}{\begin{equation}}
\newcommand{\eeq}{\end{equation}}
\newcommand{\beqa}{\begin{eqnarray}}
\newcommand{\eeqa}{\end{eqnarray}}
\newcommand{\bc}{\begin{center}}
\newcommand{\ec}{\end{center}}

\newcommand{\ba}{\begin{array}}
\newcommand{\ea}{\end{array}}
\newcommand{\pa}{\partial}

\begin{document}

	\title{{\bf{On the equivalence among stress tensors in a gauge-fluid system }}}
	\author{  {\bf {\normalsize Arpan Krishna Mitra$^1$}$
			$\thanks{E-mail: arpan@bose.res.in}}$~$
	{\bf {\normalsize Rabin Banerjee$^1$}$
			$\thanks{E-mail: rabin@bose.res.in}},$~$
		{\bf {\normalsize Subir Ghosh$^2$}$
			$\thanks{E-mail: subir\_ ghosh2@rediffmail.com}},$~$
		\\{\normalsize $^1$S. N. Bose National Centre for Basic Sciences,}
		\\{\normalsize JD Block, Sector III, Salt Lake, Kolkata-700098, India}
		\\\\
		{\normalsize $^2$Indian Statistical Institute}
		\\{\normalsize  203, Barrackpore Trunk Road, Kolkata 700108, India}
		\\[0.3cm]
	}
	\date{}
	\maketitle
	
\section*{Abstract:}  In this paper we bring out the subtleties involved in the study of a first order relativistic field theory with auxiliary field variables playing an essential role. In particular we discuss the  nonisentropic Eulerian (or Hamiltonian) fluid model. Interactions are introduced by coupling the fluid to a {\it dynamical} Maxwell ($U(1)$) gauge field. This dynamical nature of the gauge field is crucial in showing the equivalence, on the physical subspace, of the stress tensor derived from two definitions, {\it{ie.}} the canonical (Noether) one and the symmetric one. In the conventional equal-time formalism, we have shown that the generators of the spacetime transformations obtained from these two definitions agree, modulo the Gauss constraint. This equivalence in the physical sector has been achieved only because of the dynamical nature of the gauge fields. Subsequently we have explicitly demonstrated the validity of the Schwinger condition. A detailed analysis of the model in lightcone formalism has also been done where several interesting features are revealed.

{\bf{Keywords:}}	 Gauge fluid, canonical stress tensor, symmetric stress tensor, equivalence, Schwinger condition.

{\bf{PACS:}} 47.10.A, 47.10.Df.
		
		\section {Introduction}
		
	Relativistic field theories are the pillars of modern theoretical physics. It is needless to say that various features of conventional forms of field theories, ( meaning that the free theories are quadratic in nature), are well documented. 	Generically the equal-time formalism has been adopted in most cases. However, not much work has been done on an alternative form of field theory: first order form of field theory with auxiliary field variables. A very important example of this type of model is the Eulerian (or Hamiltonian) form of classical fluid (for a review from modern field theory perspective and early references see \cite{jac}). Although originally conceived as a non-relativistic theory, the relativistic generalization of hydrodynamics has been formulated in recent years\cite{lan}. However, in the Lagrangian framework, the structure of the relativistic fluid model is  qualitatively distinct from the conventional relativistic field theories, the major differences being that the former is first order in fields and depends crucially on auxiliary field variables whereas the latter are generally quadratic in fields and do not require any auxiliary variables. 
	
	Because of the unconventional structure of relativistic hydrodynamics, it is pertinent to ask how does it compare with conventional field theories, that is whether it satisfies some of the fundamental properties of the latter. In the present work we concentrate on possibly the most important object of a relativistic field theory, the energy momentum tensor. Apart from the physically relevant energy-momentum conservation principles, the tensor components act as generators of spacetime transformations that reflect the spacetime symmetries. Furthermore consistency of the relativistic model depends on the validity of the Schwinger condition \cite{nger, nge1} that is a local property and is stronger than the total energy-momentum conservation principle. The latter appears as an integrated version of the local Schwinger condition. Another point to note is that earlier works have mostly exploited the equal-time framework whereas in recent times lightcone framework is also being used in various contexts \cite{ranga, sb}. In this paper all these issues have been addressed in the context of relativistic fluid model in interaction with  Maxwell electrodynamics.

	After the above general remarks let us now elaborate in more detail our motivation and the analysis presented here.	
Hydrodynamics is one of the earliest developed  applied sciences \cite{lan} but in recent years, especially after the advent of AdS/CFT  and subsequent fluid/gravity correspondence\cite{ranga, sb}, its relevance is being appreciated in theoretical physics. From a modern high energy physics perspective, the canonical theory for relativistic perfect isentropic fluids  was developed in \cite{jac}, with special emphasis on symmetry aspects of the theory. Indeed, the classical version of ideal fluid theory is a conformal field theory and this property can be exploited in AdS/CFT  correspondence. On the other hand and more interestingly, exploiting the  fluid/gravity correspondence there is hope of  deriving a theory of non-ideal fluid and even fluid in the presence of turbulence, based on first principles. This is because, the non-ideal fluid, being a strongly coupled one, can be dual to a weakly coupled gravity theory, again thanks to AdS/CFT correspondence. The role of symmetries and their implications in fluid systems is quite crucial in this set up.

The present work is a generalization of our earlier work \cite{arpan} where we  presented a systematic and detailed analysis of an ideal relativistic isentropic fluid interacting with an external gauge field in the Hamiltonian framework. There are two extensions. First, we now consider a non-isentropic fluid. Secondly, and more importantly, the present work  deals with the full interacting theory where the gauge field is also dynamical. This additional input yields new interesting results and puts the interacting fluid model in a clearer perspective. It should be emphasized that our formalism is different from the existing works on fluid in the presence of electromagnetic interactions \cite{sc, sc1, holm, cpt, jh} that are essentially Hamiltonian in nature and do not provide a Lagrangian scenario. Moreover, the presentation in light cone coordinates is new.

 The lightcone analysis has been provided in detail, primarily because of its role in topical concepts of non-relativistic AdS/CFT and holography \cite{son} and also because of the non-trivial theoretical aspects of a relativistic theory itself in  lightcone framework. We have compared and contrasted the results with our previous  observations in \cite{arpan} that dealt with non-interacting fluid in lightcone. We have discussed in detail the question of  validity of the Schwinger condition \cite{nger}, a hallmark of a consistent relativistic theory, in the present fluid-gauge model.

As we have discussed in our earlier work \cite{arpan}, there are two distinct forms of  the stress tensor based on  two conventional definitions. The canonical $T_{\mu\nu}$ is obtained via Noether prescription and the symmetric  $\Theta_{\mu\nu}$ is obtained by metric variation. For the free theory both definitions agree. However, in the presence of interaction, $T_{\mu\nu}$ and  $\Theta_{\mu\nu}$ do not match.  Here we have been able to demonstrate that there is no inconsistency regarding this mismatch. The point is that the physically relevant quantities are the integrated versions of $T_{\mu\nu}$(or $\Theta_{\mu\nu}$) which define the various space time generators. Interestingly, the integrated versions of $T_{\mu\nu}$ and $\Theta_{\mu\nu}$ agree, modulo terms proportional to the Gauss constraint.  Hence, in the physical subspace, the two definitions of the generators agree. Indeed this has been possible only because of the dynamical nature of the gauge field which brings about new constraints in the theory, in particular the Gauss law. This constraint, incidentally, did not appear for non-dynamical gauge fields.

The paper is organized as follows: In section 2 the relativistic fluid model interacting with a dynamical $U(1)$ gauge field is formulated in terms of Clebsch variables \cite{cl,kara, lin, car,} in equal-time coordinate. The stress tensors obtained by Noether's prescription and metric variation are shown to be conserved. Although their local structures differ, their integrated versions, defining the space-time generators, are shown to be gauge equivalent.  The    Schwinger condition is also verified in equal time coordinate. In section 4 the lightcone analysis is performed. The first interesting observation is that even though the symplectic structure in the gauge sector is first order in nature, the algebra between the gauge field variables cannot be simply read off. We have employed Dirac's \cite{dir, hrt} scheme of Hamiltonian constraint analysis to derive the algebra. 
  The transformation laws and conservation principles involving the stress tensor in  light-cone formalism are studied.  The paper ends with our conclusions and future prospects in section 4.

\section{Relativistic, nonisentropic fluid mechanics in equal-time coordinates}
Let us quickly recapitulate the free fluid field theory in Eulerian approach. Construction of the fluid Lagrangian requires the introduction of Clebsch variables \cite{cl, kara, lin, car} $ \theta , \alpha , \beta , \gamma , S $, that appears in the fluid Lagrangian 
 \begin{equation}
 	\label{a}
 	{\cal{L}}=-\eta^{\mu\nu}j_{\mu}a_{\nu}-f;~~ ~ \eta^{\mu\nu}=diag(1,-1,-1,-1)
 \end{equation} in the following combination\cite{ arp1, kul, sel, sarl},
\begin{equation}
	\label{b}
	a_{\mu}=\partial_{\mu}\theta+
	\alpha\partial_{\mu}\beta +\gamma \pa_{\mu}S .
\end{equation}
We identify $S$ as the entropy.
The generalized scalar potential function $f(\sqrt{j^{\mu}j_{\mu}})$ dictates the dynamics. 
In the Eulerian description of relativistic fluid the dynamical variables are the matter density $j^0 $ and the currents $j^i, i=1,2,3$ that satisfy the conservation law,
\begin{equation}
\label{e0}
\partial_\mu j^\mu =0.
\end{equation}
From the expanded form of the Lagrangian (\ref{a}), (with $j^{\mu}j_{\mu}=n^{2}$, a relativistic scalar, and identifying  $\rho=j^{0}$ as the density), 
\begin{equation}
	\label{c}
	{\cal{L}}= -\rho\partial_{0}\theta-j^{i}\partial_{i}\theta-\rho\alpha\partial_{0}\beta-j^{i}\alpha\partial_{i}\beta-\rho\gamma\pa_{0}S - j^{i}\gamma\pa_{i}S-f(n),
\end{equation}
it is straightforward to show that the current conservation law (\ref{a}) follows from the $\theta$-equation of motion. 

Let us now posit the  relativistic version   of a fully interacting model of a fluid and a dynamical $U(1)$ gauge field as,
\begin{equation}
\label{ne1}
{\cal{L}}=-\eta^{\mu\nu}j_{\mu}(a_{\nu}-A_{\nu})-f-\frac{1}{4}F^{\mu\nu}F_{\mu\nu}.
\end{equation}
where $F_{\mu\nu }=\partial _\mu A_\nu-\partial _\nu A_\mu $ is the electromagnetic field strength. This ia a natural extension of our previous work \cite{arpan} where we considered an isentropic fluid (without the entropy term) and treated the gauge field as external.

Variations of the dynamical variables  $ \alpha, \beta,\gamma, S, \rho(=j^{0}),j^{\mu}, A_{\mu} $  yield the equations of motion,
\begin{equation}
\label{ine1}
j^\mu\partial _\mu\alpha =0,
\end{equation}
\begin{equation}
\label{ine2}
j^\mu\partial _\mu\beta =0,
\end{equation}
\begin{equation}
\label{ins}
j^\mu\partial _\mu S=0
\end{equation}
\begin{equation}
\label{ing}
j^\mu\partial _\mu \gamma=0
\end{equation}
\begin{equation}
\label{ine3}
\dot{\theta}+ \alpha\dot{\beta}+\gamma\dot{S}+\dfrac{\rho}{n}{f^{\prime }(n)}=0.
\end{equation}
\begin{equation}
\label{inr}
j_{\mu}=-\frac{n}{f^{\prime}(n)}(a_{\mu}-A_{\mu})=-\frac{n}{f^{\prime}(n)}(\partial_{\mu}\theta + \alpha \partial_{\mu}\beta +\gamma \pa_{\mu}S-A_{\mu}).
\end{equation}
\begin{equation}
\label{ins}
j_{\beta}=-\partial^{\alpha}F_{\alpha \beta}
\end{equation}
It is easy to see that current conservation (\ref{e0}) also follows from (\ref{ins}). Due to the presence of this conservation, the action corresponding to (\ref{ne1}) is invariant under the gauge transformation,
\begin{equation}
\label{gt} A_\mu \rightarrow A_\mu +\partial_\mu \Lambda, \end{equation}
exactly as happens in electrodynamics. This similarity persists further by noting that, as in electrodynamics, there occurs a Gauss constraint which is given by the time component of (\ref{ins}) ,
\begin{equation}
\partial_i\pi_i-j_0=\partial_i\pi_i-\rho =0,
\label{gau}
\end{equation}
where $\pi_i=\frac{\partial {\cal{L}}}{\partial \dot{A}^i}=F_{i0}$ is the momentum conjugate to $A^i$. The Gauss constraint is the generator of the gauge transformation (\ref{gt}) and defines the physical subspace as 
\begin{equation}
\label{yomo}
(\partial_i\pi_i-\rho )\mid \Psi>_{Physical} =0.
\end{equation}
From (\ref{c}) we can identify three  independent canonical pairs  $(\rho , \theta ) $, $(\alpha \rho, \beta )$ and $(\rho \gamma, S)$.
The fundamental brackets, compatible with the above canonical pairs, follow from the symplectic structure,
   \begin{equation}
   \nonumber
   \lbrace\rho(x),\theta(y)\rbrace=\delta{\bf{(x-y)}},~ \lbrace\alpha(x),\theta(y)\rbrace=-\frac{\alpha}{\rho}\delta{\bf{(x-y)}},~\lbrace\alpha(x),\beta(y)\rbrace=\dfrac{\delta{\bf{(x-y)}}}{\rho};
   \end{equation}
   \begin{equation}
   \label{f}
   \lbrace\gamma(x),S(y)\rbrace=\dfrac{\delta{\bf{(x-y)}}}{\rho},~~\{\gamma(x), \theta(y)\}=-\frac{\gamma}{\rho}\delta{\bf{(x-y)}}.
   \end{equation}

 All other brackets are vanishing. It is important to note that the apparent singularity in the above symplectic structure  for $\rho \rightarrow 0$ does not create any problem simply because this limit is unphysical since the kinetic part of Lagrangian in (\ref{c}) completely disappears for $\rho = 0$.
 
We now concentrate on the structure of the energy-momentum tensor. Conventionally there are two parallel definitions. One of these is the symmetric energy-momentum tensor, 
\begin{equation}
\label{s}
\Theta_{\mu \nu}=-\frac{2}{\sqrt{-g}}\frac{\partial S}{\partial g^{\mu\nu }},
\end{equation} 
that is obtained by generalizing (\ref{ne1}) to a curved spacetime which amounts to replacing $\eta^{\mu\nu}$ by $g^{\mu\nu}$, varying $g^{\mu\nu}$ and finally reverting back to flat spacetime with the replacement of  $g^{\mu\nu}$ by $\eta^{\mu\nu}$ in (\ref{s}).
\pagebreak\\

 On the other hand, the canonical energy-momentum tensor is obtained via Noether prescription,
\begin{equation}
\label{newnoe}
T_{\mu\nu}=\frac{\partial {\cal{L}}}{\partial(\partial^{\mu}\theta)}\partial_{\nu}\theta + \frac{\partial {\cal{L}}}{\partial(\partial^{\mu}\beta)}\partial_{\nu}\beta +\frac{\partial {\cal{L}}}{\partial(\partial^{\mu}\alpha)}\partial_{\nu}\alpha +\frac{\partial {\cal{L}}}{\partial(\partial^{\mu}\rho)}\partial_{\nu}\rho +\frac{\partial {\cal{L}}}{\partial(\partial^{\mu} S)}\partial_{\nu} S +\frac{\partial {\cal{L}}}{\partial(\partial^{\mu} A^{\lambda})}\partial_{\nu}A^{\lambda} - \eta_{\mu\nu}{\cal{L}}$$$$
\end{equation}
Both the definitions have their utilities. $T_{\mu \nu}$ is designed to manifestly generate correct space-time transformations of the field variables but it is not symmetric (and can be improved by Belinfante prescription) whereas $\Theta_{\mu \nu}$ is manifestly symmetric but its ability to generate appropriate 
space time transformation is not transparent. In simple cases these expressions agree as is natural but there are subtleties involved in the fluid system under consideration. We emphasize that these issues have not been studied so far but become crucial for the consistency of the fluid model.

In our interacting fluid model, the canonical energy-momentum tensor is given by \eqref{newnoe},
\begin{equation}
T_{\mu\nu}=-j_{\mu}(\partial_{\nu}\theta+\alpha \partial_{\nu}\beta + \gamma \partial_{\nu}S ) -F_{\mu \sigma} \partial_{\nu}A^{\sigma}-\eta_{\mu\nu}{\cal{L}}.
\end{equation}
This tensor is conserved. To show this explicitly, we exploit current conservation  and other equations of motion to find,
\begin{equation}
\nonumber
\pa^{\mu}T_{\mu\nu}=\pa^{\mu}\{-j_{\mu}(\pa_{\nu}\theta+\alpha\pa_{nu}\beta +\gamma\pa_{\nu}S)-F_{\mu\sigma}\pa_{\nu}A^{\sigma}-\eta_{\mu\nu}{\cal{L}}\}
\end{equation}
\begin{equation}
\label{yoooo}
=(\pa_{\nu}j^{\mu})(a_{\mu}-A_{\mu})+\pa_{\nu}f(n)=0
\end{equation}
where the final step is obtained on using \eqref{inr}.

On the other hand the symmetric energy momentum tensor is derived from (\ref{s})  as,
\begin{equation}
\label{cor9}
\Theta_{\mu\nu}=-\eta_{\mu \nu}{\cal{L}}+\frac{j_{\mu} j_{\nu}}{n}f'-F^{\beta}~ _{\nu}F_{\beta \mu}.
\end{equation}
This is also conserved by applyig the various equations of motion,
\begin{equation}
\nonumber
\pa^{\mu}\Theta_{\mu\nu}=0.
\end{equation}
Now $\Theta_{\mu\nu}$ produces the Hamiltonian 
\begin{eqnarray}
\Theta_{00}=-{\cal{L}}+\frac{j_{0}j_{0}}{n}f'-F^{j}~_{0}F_{j0}\\
=-{\cal{L}}-\rho(\partial_{0}\theta+\alpha \partial_{0}\beta + \gamma \partial_{0}S)+\rho A_{0}-F^{j}~_{0}F_{j0}.
\end{eqnarray}
Also $T_{\mu\nu}$ in (\ref{newnoe}) gives rise to canonical Hamiltonian,
\begin{equation}
T_{00}=-\rho(\partial_{0}\theta+\alpha \partial_{0}\beta +\gamma \partial_{0}S)-F_{0\sigma}\partial_{0}A^{\sigma}-{\cal{L}}.
\end{equation}
Let us compute the difference between two Hamiltonian densities,
\begin{equation}
T_{00}-\Theta_{00}=-F_{0i}\partial_{0}A^{i}+F^{j}~_{0}F_{j0}-\rho A_{0}
\end{equation}
\begin{equation}
=-\pi_{i}\partial_{0}A_{i}-\pi_{i}^{2}-\rho A_{0}=-\pi_{i}(\partial_{i}A_{0}-\pi_{i})-\pi_{i}^{2}-\rho A_{0}
\end{equation}
\begin{equation}
=-\pi_{i}\partial_{i}A_{0}-\rho A_{0}.
\end{equation}
which is obviously nonvanishing. However the physically relevant object is the integrated version which corresponds to the hamiltonian. This difference in the hamiltonians is found to be,
\begin{eqnarray}
\label{ham}
\int~d^3x (T_{00}-\Theta_{00})=-\int~d^3x (\pi_{i}\partial_{i}A_{0}+\rho A_{0})\\
=\int ~d^3x A_{0}(\partial_{i}\pi_{i}-\rho).
\end{eqnarray}
which is proportional to the Gauss law. Hence, on the physical surface \eqref{yoooo} these two expressions are identical. This result should be contrasted with our previous observation in \cite{arpan} where also this mismatch was noted but since the gauge field was not dynamical there was no Gauss law and this mismatch persisted. Clearly the kinetic part of the gauge field, which is the Maxwell term, rounds off the theory nicely. But it is still necessary to check if the same property holds for the other important components of the stress tensor. 

Let us consider the momentum density. The relevant expressions are,
\begin{eqnarray}
T_{0i}=-\rho(\partial_{i}\theta+
\alpha\partial_{i}\beta + \gamma \partial_{i}S)-F_{0j}\partial_{i}A^{j} 
=-\rho(\partial_{i}\theta+\alpha\partial_{i}\beta+
\gamma \partial_{i}S)-\pi_{j}\partial_{i}A^{j},
\end{eqnarray}
\begin{eqnarray}
\label{momd}
\Theta_{0i}=\frac{j_{0}j_{i}}{n}f'-F^{\beta}~_{i}F_{\beta 0}
=-\rho(\partial_{i}\theta+\alpha\partial_{i}\beta + \gamma \partial_{i}S)+\rho A_{i}-F^{k}~_{i}F_{k0},
\end{eqnarray}
with the difference
\begin{eqnarray}
\label{diffm}
T_{0i}-\Theta_{0i}=-\rho A_{i}-\pi_{j}\partial_{i}A_{j}+\pi_{j}(\partial_{i}A_{j}-\partial_{j}A_{i})
=-(\rho A_{i}+\pi_{j}\partial_{j}A_{i}).
\end{eqnarray}
Once again integration of the above result yields  
\begin{equation}
\label{mom}
\int~d^3x (T_{0i}-\Theta_{0i})= \int~d^3x A_{i}(\partial_{i}\pi_{i}-\rho),
\end{equation}
indicating that the total momenta in the two definitions are equal modulo the first class (Gauss) constraint. Exploiting the covariant notation, the
combination of (\ref{ham}) and (\ref{mom}) is written in a   compact form,
\begin{equation}
\label{uvl}
\int~d^3x (T_{0\mu}-\Theta_{0\mu})= \int~d^3x A_{\mu}(\partial_{i}\pi_{i}-\rho),
\end{equation}
which vanishes on the physical subspace.

It is possible to continue  this analysis for the angular momentum operator. From Noether's definition, this is given by,
\begin{equation}
\label{angn}
M_{ij}^{N}=\int (x_i T_{0j}-x_{j}T_{0i}-\frac{\pa {\cal{L}}}{\pa\dot{A}^{\lambda}}\Sigma^{\lambda\sigma}_{ij}
A_{\sigma})d^{3}x
\end{equation}
where the spin tensor is defined as, 
\begin{equation}
\label{spin}
\Sigma^{\lambda\sigma}_{\alpha\beta}= g^{\lambda}_{\alpha}~g^{\sigma}_{\beta}-g^{\lambda}_{\beta}~g^{\sigma}_{\alpha}.
\end{equation}

We therefore obtain,$$M_{ij}^{N}=\int (x_i T_{0j}-x_{j}T_{0i}-\pi_{i}A_{j}+\pi_{j}A_{i}) d^3 x$$

The angular momentum, following from the symmetric tensor \eqref{s}, is given by $$M_{ij}^{S}=\int (x_i \Theta_{0j}-x_{j}\Theta_{0i})d^3 x$$

Using \eqref{momd} and \eqref{diffm} it is seen that the difference between these expressions vanishes, modulo terms proportional to the Gauss constraint,
\begin{equation}
\label{angd}
M_{ij}^{N}-M_{ij}^{S}=\int d^3 x (x_{i}A_{j}-x_{j}A_{i})(\pa_{k}\pi_{k}-\rho).
\end{equation} 

Thus on the physical subspace, the expressions for angular momenta are identical, as happened for the space-time translation generators discussed earlier.

Similarly, the difference in the structures of the boost generators can also   be discussed. From Noether's definition, the boost is given by,
\begin{equation}
\label{boo1}
M_{0i}^{N}=\int (x_0 T_{0i}-x_{i}T_{00}-\frac{\pa {\cal{L}}}{\pa\dot{A}^{\lambda}}\Sigma^{\lambda\sigma}_{0i}.
~A_{\sigma})d^{3}x
\end{equation}
From \eqref{spin} it follows,
\begin{equation}
\label{boon}
M_{0i}^{N}=\int (x_0 T_{0i}-x_{i}T_{00}-\pi_{i}A_{0})d^{3}x.
\end{equation}

On the other hand, the  definition of boost following from the symmetric tensor \eqref{s} is,
 \begin{equation}
\label{boo2}
M_{0i}^{N}=\int (x_0 \Theta_{0i}-x_{i}\Theta_{00})d^{3}x .
\end{equation}
Once again the difference is just proportional to the Gauss constraint, 
\begin{equation}
\label{bood}
M_{0i}^{N}-M_{0i}^{S}=\int d^3 x (x_{0}A_{i}-x_{i}A_{0})(\pa_{k}\pi_{k}-\rho)
\end{equation}
In fact \eqref{angd} and \eqref{bood} maybe combined to yield a covariant structure,
\begin{equation}
\label{cood}
M_{\mu i}^{N}-M_{\mu i}^{S}=\int d^3 x (x_{\mu}A_{i}-x_{i}A_{\mu})(\pa_{k}\pi_{k}-\rho).
\end{equation}
Indeed, the above exercise is non-trivial since it underlines the importance of introducing the Maxwell gauge field kinetic term and also establishes the spacetime symmetries of the fully interacting relativistic fluid model in a robust way. This explicit demonstration was absent in previous literatures.

{\bf{Schwinger condition:}} 

In its simplest form, the Schwinger  covariance condition
 relates the equal-time energy
density commutator to the momentum density, 
 \begin{equation}
[\Theta_{00}(x),\Theta_{00}(x')]= (\Theta_{0i}(x)+\Theta_{0i}(x'))\pa_i\delta (x-x'). 
\end{equation}
For some quantum field theoretical applications see \cite{br}, where it is referred to as Dirac-Schwinger condition \cite{ds}.) Validity of this condition  in a quantum field theory ensures that
the theory is relativistically  covariant. However, it can play an important role in  field theories even in non-relativistic scenario \cite{arpan, arp1}.

Let us now concentrate on the Schwinger condition for the present model. In our previous paper \cite{arpan} we have demonstrated the validity of the Schwinger condition for the non-interacting fluid model. The situation is more complicated here because the gauge fields being dynamical satisfy a canonical Poisson algebra $\{A_i(\bf x),\pi ^j(\bf y)\}=\delta_i^j\delta (\bf x - \bf y)$. We need to compute the following bracket,
\begin{equation}
\{\Theta_{00}(x),\Theta_{00}(y)\}=\{j^{i}(\pa_{i}\theta +\alpha\pa_{i}\beta-A_{i})+f+\frac{1}{4}F^{ij}F_{ij}+\frac{1}{2}\pi_{i}^{2}|_{x},j^{k}(\pa_{k}\theta +\alpha\pa_{k}\beta-A_{k})+f+\frac{1}{4}F^{lm}F_{lm}+\frac{1}{2}\pi_{k}^{2}|_{y}\}.
\end{equation}
After a long but straightforward calculation, we arrive at the result,
\begin{equation}
\nonumber
\{\Theta_{00}(x),\Theta_{00}(y)\}=[(-\rho(\pa_{i}\theta+\alpha\pa_{i}\beta-A_{i})+F_{ik}\pi^{k}|_{x})+(-\rho(\pa_{i}\theta+\alpha\pa_{i}\beta-A_{i})+F_{ik}\pi^{k}|_{y})]\pa_{i}^{x}\delta (x-y)
\end{equation}
\begin{equation}
=(\Theta_{0i}(x)+\Theta_{0i}(y))\pa_{i}^{x}\delta (x-y).
\end{equation}
This ensures the validity of the Schwinger condition in the fully interacting fluid-Maxwell theory.
\section{Light cone analysis}

Light-cone (or light front form of) quantization (LCQ) was introduced very early with two principal motivations: as a computational tool for bound state solutions in QCD to represent hadrons as bound states of quarks and gluons in a relativistic framework and also to utilize computers in quantum field theory calculations. (See \cite{light rev} for an early review.) In fact the convenience of LCQ was pointed out by Dirac \cite{light dir} as an alternative to equal time quantization where the lightcone coordinates  are defined as as \cite{hrt} $\lbrace x^{+}, x^{-}, \bar{x}\rbrace$,  where
\begin{equation}
x^{\pm}=\frac{1}{\sqrt 2}(x^0\pm x^3);~~\bar{x}\equiv x^a= x^1,x^2.
\end{equation}
 Here $ x^{+}$ plays the role of time and $\bar{x}$ are referred as transverse coordinates.  The non-vanishing  metric components are,
\begin{equation}
 g^{+-}=g^{-+}=1;~~g^{ab}=\delta^{ab},~ a,b=1,2.
\end{equation}
    In the context of QCD a related framework, known as  Infinite Momentum Frame,  was initiated \cite{fub, wein} to explain Bjorken scaling in scattering phenomena. The  physical meaning of this correspondence is that measurements made by an observer moving at infinite momentum is equivalent to making observations with speed being close to  the speed of light and this   corresponds to the front form where measurements are made along the front of a light wave.
 
 Coming back to recent times LCQ has generated tremendous amount of interest after the celebrated work of Son \cite{son} who formulated a model that represented the experimentally demonstrated  trapping of cold atoms  at  Feshbach resonance, thereby introducing the concept of non-relativistic holographic principle in AdS/CFT correspondence. It is important to note that in light cone variables a second order system, (in terms of time derivative), such as Klein Gordon, is changed to a first order system such as Schrodinger. But precisely this  algebraic manipulation drastically alters the Hamiltonian structure of the system because the converted first order system turns out to be a constraint system with a non-canonical symplectic structure and reduced number of degrees of freedom. We will explicitly demonstrate that there are subtleties involved in the Hamiltonian analysis since the lightcone coordinate system is qualitatively distinct from the conventional equal time coordinate framework. At this point it is worthwhile to recall our earlier work \cite{arpan}
where, for the first time, a detailed lightcone analysis of the free fluid system was performed. There \cite{arpan} it was observed that the symplectic structure in lightcone coordinate did not differ from the one in  equal time coordinate, the reason being that the free fluid model was a first order system even in  equal time coordinate. However, the difference between the two frameworks was manifest in {\it{eg}}. Schwinger condition where  the spatial coordinates, $x_-$ and transverse ones $\bar x$, were clearly separated into different sectors. In the present work, where we consider the fully interacting fluid-Maxwell theory,   the situation becomes much more serious since the Maxwell gauge sector is quadratic in nature and upon LCQ leads to complications that puts a question mark on the validity of the Schwinger condition. This is not surprising since, in the hamiltonian framework, LCQ even for a simple massless scalar theory involves subtleties and complications. However, we emphasize, that the total energy of the system remains conserved in LCQ.

\subsubsection*{{Massless scalar:}} The intricacies of LCQ can be seen in the simplest of models, that of a massless scalar field. It also helps in setting up the notation and introduce some basic formula. The Lagrangian
\begin{equation}
L=\int d^4x~\frac{1}{2}\partial_\mu \phi \partial^\mu \phi ,
\label{ll}
\end{equation}
generates the equation of motion,
\begin{equation}
\pa_\mu \pa^\mu \phi =0.
\label{msc}
\end{equation}
The same equation is recovered from the Hamiltonian
\begin{equation}
H=\int d^3x~{\cal{H}}(x)=\int d^3x~\frac{1}{2}(\pi ^2+\partial_i \phi \partial_i \phi ),
\label{h1}
\end{equation}
as  Hamilton's equation of motion where  the equal-time canonical algebra,

\begin{equation}
\label{newa}
\{\phi(\bar x),\pi(\bar y) \}=\delta(\bar x -\bar y), \{\phi(\bar x),\phi(\bar y) \}=\{\pi(\bar x),\pi(\bar y) \}=0 
\end{equation}
is used. 

It is now
 straightforward to compute the bracket between the energy densities $\{{\cal{H}}( x),{\cal{H}}(y)\}$ to yield,
\begin{equation}
\{\frac{1}{2}(\pi ^2+\partial_i \phi \partial_i \phi )(x), \frac{1}{2}(\pi ^2+\partial_i \phi \partial_i \phi )(
 y)\}=((\pi \partial_i\phi)(x)+\pi \partial_i\phi)(y))\partial_i\delta (\bar x - \bar y).
\label{ab}
\end{equation}
The above equation amounts to,
\begin{equation}
\{{\cal{H}}(x),{\cal{H}}(y)\}=({\cal{P}}_i(x)+{\cal{P}}_i(y))\partial_i\delta (\bar x - \bar y),
\end{equation}
thereby yielding the Schwinger condition where ${\cal{P}}_i=\pi\pa_{i}\phi$ is defined as the momentum density.

The same Lagrangian, now expressed  in lightcone coordinates,
\begin{equation}
{\cal{L}}=\pa_+\phi \pa_-\phi -\frac{1}{2}\pa_i\phi \pa_i\phi
\label{l1}
\end{equation}
generates the equation of motion,
\begin{equation}
2\pa_+ \pa_- \phi =\pa_i\pa_i \phi ,
\label{msc1}
\end{equation}
which is identical to (\ref{msc}). However, recovering (\ref{msc1}) in Hamiltonian formalism is more complicated since the lightcone Lagrangian (\ref{l1}) is a constraint system in Dirac's formulation of constraint dynamics. Note that the momentum, defined as $\pi =(\pa {\cal{L}})/(\pa(\pa_+\phi ))=\pa_-\phi $, does not contain  a time derivative term and hence (in Dirac's scheme \cite{dir}) is interpreted as a primary constraint,  
\begin{equation}
\Omega (x)\equiv \pi (x)-\pa_-\phi (x)\approx 0.
\label{om}
\end{equation} The $\approx$ indicates that it is a weak equality which cannot be strongly imposed. This has important implications in the computation of the Poisson algebra of any variable with $\Omega(x)$. Naively, this would vanish. However, due to the weak equality, this is no longer valid and an explicit computation is necessary. Indeed we find, 
\begin{equation}
\{\Omega (x), \Omega (y) \}=2\pa_-\delta(x_--y_-)\delta(\bar x-\bar y).
\label{0l1}
\end{equation}

Canonical symplectic structure \eqref{newa} is used to derive (\ref{0l1}). Since the constraint algebra \eqref{0l1} does not close, the constraint $\Omega(x)$ is said to be second class in the sense of Dirac \cite{dir}. This feature is typical of second order systems when expressed in lightcone variables.
The next step is to compute the Dirac brackets (denoted by a star) which are defined in terms of the Poisson brackets by,
\begin{equation}
\label{cor2}
\{A(x), B(y)\}^{*} = \{A(x), B(y)\}-\int \{A(x), \Omega(z_{1})\}\{\Omega(z_{1}), \Omega(z_{2})\}^{-1}
\{\Omega(z_{2}), B(y)\}dz_{1}dz_{2}
\end{equation}
where $\{\Omega(z_{1}), \Omega(z_{2})\}^{-1}$ is the inverse of \eqref{0l1} defined as,
\begin{equation}
\label{cor3}
\int~dy \{\Omega(x), \Omega(y)\}\{\Omega(y), \Omega(z)\}^{-1}= \delta(x-z)
\end{equation}
By introducing the sign function $\epsilon(x_{-}-y_{-})$ given by, 
\begin{equation}
\label{cor4}
\pa_{x_{-}}\epsilon(x_{-}-y_{-})=\delta(x_{-}-y_{-})
\end{equation}
it is simple to show that the inverse has the form,
\begin{equation}
\label{cor5}
\{\Omega(x), \Omega(y)\}^{-1}=\frac{1}{2}\epsilon(x_{-}-y_{-})\delta(\bar{x}-\bar{y})
\end{equation}
It is now possible to compute the Dirac brackets among the field variables,
\begin{equation}
	\label{n11}
	\{\phi(x_{+},x_{-}, \bar{x}),\phi(y_{+},y_{-}, \bar{y})\}=-\int dz_1~dz_2~\{\phi(x),\Omega(z_1)\} [\{\Omega (z_1), \Omega (z_2) \}]^{-1}\{\Omega(z_2),\phi(y)\} $$$$=	
	\frac{1}{2}\epsilon(x^{-}-y^{-})\bar{\delta}(\bar{x}-\bar{y}).
\end{equation}
The advantage of using Dirac brackets is that the second class constraints can now be strongly imposed. Thus the Dirac brackets of $\Omega(x)(\ref{om})$ with any variable vanishes, as may be easily checked.

Note the non-local nature of the lightcone symplectic structure \eqref{n11}. Together with the Hamiltonian 
\begin{equation}
H=\int dy^-d\bar y~\pa_i\phi \pa_i\phi
\label{hh1}
\end{equation}
and the algebra (\ref{n11}), we compute $\partial_+\phi$,
\begin{equation}
	\label{n13}
	\pa_{+}\phi =\pa_{\bar{x}}^{i}\int dy^{-} ~\pa_{i}\phi(x_{+},\bar{x},y^{-})\frac{1}{2}\epsilon(x^{-}-y^{-}).
\end{equation}
Furthermore, on differentiating both sides by  $\partial_-$ the non-locality is removed and one recovers the correct equation of motion \eqref{msc1}.
This clearly underlines the fact that lightcone framework, while reproducing the equal time equation of motion \eqref{msc}, is qualitatively distinct from the equal time framework with a reduced number of degrees of freedom due to the constraint. The original second order system is converted to first order.

To derive the energy conservation principle,  from the covariant form of the symmetric energy momentum tensor for the massless scalar,
\begin{equation}
\label{emtenm}
\Theta^{\mu\nu}=\pa^{\mu}\phi\pa^{\nu}\phi-
{\cal{L}}\eta^{\mu\nu}
\end{equation}
we first write down the different lightcone components as,
\begin{equation}
\label{llc}
 \Theta^{--}=(\pa_{+}\phi)^{2},~~
 \Theta^{i-}=-\pa_{i}\phi\pa_{+}\phi, ~~{\cal{H}}\equiv\Theta^{+-}=\pa_{i}\phi\pa_{i}\phi 
\end{equation}
where $\Theta^{+-}$ is identified with the 
Hamiltonian density ${\cal{H}}.$

Let us now calculate the time derivative of $\Theta^{+-}$,
\begin{equation}
\pa_{+}\Theta^{+-}=\{{\cal{H}}(x), H\}=\{\frac{1}{2}\pa_i\phi (x) \pa_i\phi (x) , \int dy^-d\bar y~\frac{1}{2}\pa_i\phi \pa_i\phi\}.
\label{dbrac}
\end{equation}
Using the algebra (\ref{n11}) we find
\begin{eqnarray}\nonumber\\
\{\frac{1}{2}\partial_i\phi(x)\partial_i\phi(x), \int~dy^-d\bar{y} ~\frac{1}{2}\partial_j\phi(y)\partial_j\phi(y)\}
\nonumber\\
=\frac{1}{2}(\pa_{i}\phi)(x)\pa_{i}^{x}\int~dy^-d\bar{y} (\pa_{j}\pa_{j}\phi){(y)}\epsilon(x^{-}-y^{-})\delta^{2}(\bar{x}-\bar{y}).\nonumber
\end{eqnarray}
Exploiting the lightcone equation of motion \eqref{msc1} we obtain,
\begin{equation}
\label{uuooo}
\pa_{+}\Theta^{+-}=\pa_{i}\phi\pa_{i}\pa_{+}\phi=-\partial_-[\partial_+\phi(x)^2]+\partial_i[ \partial_+\phi(x)(\partial_i\phi(x)],
\end{equation}
where the final step is obtained after a simple algebra and reusing \eqref{msc1}. The factors in parenthesis are identified with $\Theta^{--}$ and $\Theta^{i-}$, respectively, as seen from (\ref{llc}). Then equation (\ref{uuooo}) is further rewritten as,
\begin{equation}
\label{zero}
\pa_{\mu}\Theta^{\mu -}=\pa_{+}\Theta^{+-}+\pa_{-}\Theta^{--}+
\pa_{i}\Theta^{i-}=0
\end{equation}

This validates  the conservation of energy.
Likewise the other components of $\pa_{\mu}\Theta^{\mu\nu}=0$ can be shown to hold.

\subsubsection*{{Interacting fluid:}}  Returning to the interacting fluid model, let us rewrite the Lagrangian (\ref{ne1}) in lightcone variables,
\begin{eqnarray}
\label{lcl}
\nonumber
{\cal{L}}= -j^{+}(\pa_{+}\theta +\alpha\pa_{+}\beta - A_{+})- j^{-}(\pa_{-}\theta +\alpha\pa_{-}\beta - A_{-})-j^{i}(\pa_{i}\theta +\alpha\pa_{i}\beta - A_{i})- f\\- \frac{1}{4}(2F^{+-}F_{+-}+ 2F^{+i}F_{+i}+2F^{-i}F_{-i}+f^{ij}F_{ij}).
\end{eqnarray}

with,
\begin{equation}
\label{cor15}
a_{\mu}= \partial_{\mu}\theta+
	\alpha\partial_{\mu}\beta .
\end{equation}
 
Interestingly the symplectic structure in the fluid sector remains essentially unaffected since it was already in a first order form (in equal time framework in (\ref{ne1}). Hence the previous fluid algebra (\ref{f}) suffices. But the constraint structure in the gauge sector is much more involved.
 The conjugate momenta are,
\begin{equation}
\label{cor6}
\pi^{\mu}=\frac{\partial{\cal{L}}}{\partial ({\pa_+A^\mu})}=F^{\mu +}
\end{equation}

  Only $\pi^{-}$ is a true momentum since it involves a time derivative. The other components have to be interpreted as primary constraints,
\begin{eqnarray}
\label{lcc}
\Omega_{1}=\pi^{+}\approx 0,~~\chi_a=\pi^{a}-F^{a +}\approx 0,~~a=1,2.
\end{eqnarray}
The constraint sector $\chi_{a}$ does not close, $$\{\chi_{a}(x), \chi_{b}(y)\}=\pa^{x_{-}}\delta(x-y)\delta_{ab}$$ and hence is second class. On the other hand $\Omega_{1}$ closes, 

$$\{\Omega_{1}(x),\Omega_{1}(y)\}= \{\Omega_{1}(x),\chi_{a}(y)\}=0$$
and hence yields the first class sector. To find the secondary constraints, if any, we have to check the time conservation of $\Omega_{1}(x)$. To do this the canonical hamiltonian has to be found.

This is obtained following the conventional definition
\begin{eqnarray}
\label{lch}
\int{\cal{H}}=\int (\pi^{+}\pa_{+}A_{+} +\pi^{-}\pa_{+}A_{-}+ \pi^{a}\pa_{+}A_{a}- {\cal{L}}) =\int [\frac{1}{2}(\pi^{-})^{2}+\frac{1}{2}F^{12}F_{12}+(\pi^{-}\pa_{-}+\pi^{i}\pa_{i}-j^{+})A^{-}].
\end{eqnarray} 
Calculating the Poisson bracket of $\Omega_{1}$ with the Hamiltonian yields a secondary constraint $\Omega_{2}(x)$,
\begin{eqnarray}
\label{lcg}
\Omega_{2}(x)=\{\Omega_{1}(x), \int {\cal{H}}\} =\{\pi^{+}(x),\int dy^-d\bar y~ {\cal{H}}(y)\}=\pa_{i}\pi^{i}(x)+\pa_{-}\pi^{-}(x)+j^{+}(x)\approx 0.
\end{eqnarray}

This is just the time $(+)$ component of the equation of motion \eqref{ins}. It is referred as the Gauss constraint since it is the analogue of the Gauss law in pure electrodynamics $(\nabla . {\pi}=\nabla . {E}=0).$ No further constraint is generated by $\Omega_{2}(x)$ since, $$\{\Omega_{2}(x), \int {\cal{H}}\}=0.$$   Now $\Omega_{1}, \Omega_{2}$ constitute a set of first class constraints indicating a gauge symmetry whereas $\chi_a $, as already stated, turn out to be a second class set of constraints. Using this set of second class constraint and following the previous analysis, the nonvanishing Dirac brackets turn out to be ,
\begin{equation}
\label{lcd}
\{\pi^{-}(x), A_{i}(y)\}=\frac{1}{4}\pa_{i}^{x}\epsilon(x^{-}-y^{-})\delta^{2}(\bar{x}-\bar{y}),~
\{\pi^{-}(x), \pi^{-}(y)\}=-\frac{1}{4}\nabla^{2}(x)\epsilon(x^{-}-y^{-})\delta^{2}(\bar{x}-\bar{y}),$$$$
\{A^{i}(x), A^{j}(y)\}=\frac{1}{4}\epsilon(x^{-}-y^{-})\delta^{2}(\bar{x}-\bar{y})\delta_{ij}.
\end{equation}
For computational details the reader is encouraged to consult \cite{hrt}.

First of all we ensure that our earlier observation regarding the equality of the two definitions of the (integrated) energy momentum tensor modulo Gauss constraint remains valid in lightcone. For this we explicitly write down the different components of $T^{\mu\nu}$ and $\Theta ^{\mu\nu}$. Following the Noether's prescription we have,
 \begin{equation}
 \label{cor10}
 T_{\mu\nu}=\frac{\partial {\cal{L}}}{\partial(\partial^{\mu}\theta)}\partial_{\nu}\theta + \frac{\partial {\cal{L}}}{\partial(\partial^{\mu}\beta)}\partial_{\nu}\beta +\frac{\partial {\cal{L}}}{\partial(\partial^{\mu}\alpha)}\partial_{\nu}\alpha +\frac{\partial {\cal{L}}}{\partial(\partial^{\mu}j^{+})}\partial_{\nu}j^{+} +\frac{\partial {\cal{L}}}{\partial(\partial^{\mu} A^{\lambda})}\partial_{\nu}A^{\lambda} - \eta_{\mu\nu}{\cal{L}}$$$$
 \end{equation}
 Using \eqref{lcl} we get after simplification,
 \begin{equation}
 \label{cor11}
  T^{\mu\nu}= -j^{\mu}(\pa^{\nu}\theta +\alpha\pa^{\nu}\beta )-F^{\mu\sigma}\pa^{\nu}A_{\sigma} -\eta^{\mu\nu}{\cal{L}}.
 \end{equation}
 Now, it is straightforward to get different components of $T^{\mu\nu}$ explicitly,
\begin{eqnarray}
\label{calc}
\nonumber
T^{+-}=\frac{1}{2}(\pi ^{-})^{2}+\frac{1}{2}F_{12}F^{12}+(\pi^{-}\pa_{-}+\pi^{i}\pa_{i})
A^{-}+f\\
\nonumber +j^{-}(\pa_{-}\theta+\alpha\pa_{-}\beta)+j^{i}(\pa_{i}\theta+\alpha\pa_{i}\beta)-j^{\mu}A_{\mu},\\~~T^{+i}=-j^{+}(\pa^{i}\theta+\alpha\pa^{i}\beta)+F^{-+}\pa^{i}
A_{-}-F^{+j}\pa^{i}A_{j},
\end{eqnarray}
While the the second relation follows trivially from \eqref{cor11}, some algebra is needed to obtain the first relation.

Now, to give the components of $\Theta^{\mu\nu}$  we start from \eqref{cor9}. The explicit calculation of $\Theta^{+i}$ gives us,
\begin{equation}
\label{cor12}
\Theta^{+i}=\frac{j^{+}j^{i}}{n}f' + F^{\beta +}F_{\beta i}
\end{equation} 
which on use of \eqref{inr} produces,
\begin{equation}
\Theta^{+i}=-j^{+}(\pa^{i}\theta+\alpha\pa^{i}\beta-A^{i})+F^{-+}F_{-}^{i}+F_{ij}\pi_{j}.
\end{equation}
Similarly we can easily compute,
\begin{eqnarray}
\label{sylc}
\nonumber
\Theta^{+-}=\frac{1}{2}(\pi ^{-})^{2}+\frac{1}{2}F_{12}F^{12}+j^{-}(\pa_{-}\theta+\alpha\pa_{-}\beta -A_{-})+j^{i}(\pa_{i}\theta+\alpha\pa_{i}\beta -A_{i})+f,\\
\end{eqnarray}
It is easy to check that the following relations hold:
\begin{equation}
\label{lcsy}
\int  dy^-d\bar y~ (T^{+-}-\Theta^{+-})=-\int  dy^-d\bar y~ (\pa_{i}\pi^{i}+\pa_{-}\pi^{-}+j^{+})A^{-},
\end{equation}
\begin{equation}
\label{lcsc}
\int  dy^-d\bar y~(T^{+i}-\Theta^{+i})=-\int  dy^-d\bar y~ (\pa_{i}\pi^{i}+\pa_{-}\pi^{-}+j^{+})A^{i}.
\end{equation}
Integrated forms of the  canonical and symmetric structures of the energy momentum tensor differ by a term proportional to the Gauss constraint \eqref{lcg} and hence are equal in the physical subspace in lightcone coordinates. Equations (\ref{lcsy}, \ref{lcsc}) are the light cone analogues of the equal time relations given in \eqref{uvl}.

The integrated energy-momentum tensors execute the spacetime translations. This naturally leads to the question as to what happens to rest of the spacetime translations that is rotation. Since the derivation is somewhat tricky we give the details below.

Once again the basic distinction between the spatial coordinates $y_-$ and $\bar y$ comes in to play and we need to perform the calculations individually.  First of all we provide integrated  expressions for the $12$-component of angular momentum, $M_{12}$, in the two  definitions, derived respectively from   \eqref{angn} and  \eqref{s},
\begin{equation}
	\label{lcan}
	M_{12}^{N}=\int dx^{-}d\bar{x} \left\{ (x_{1}T_{-2}-x_{2}T_{-1})-(\pi_{1}A_{2}-\pi_{2}A_{1})\right\},
\end{equation}
\begin{equation}
	\label{lcas}
	M_{12}^{S}=\int dx^{-}d\bar{x} \{ (x_{1}\Theta_{-2}-x_{2}\Theta_{-1})\}.
\end{equation}
Difference between these two expressions is computed below,
\begin{equation}
\nonumber
	M_{12}^{N}-M_{12}^{S}=\int dx^{-}d\bar{x} ~\{ x_{1}(T_{-2}-\Theta_{-2})-x_{2}(T_{-1}-\Theta_{-1})-(\pi_{1}A_{2}-\pi_{2}A_{1})\}
	\end{equation}
\begin{equation}
\nonumber
=\int dx^{-}d\bar{x}~\{ x_{1}(-j^{+}A_{2}+\pi^{-}\pa_{-}A_{2}+\pi^{i}\pa_{i}A_{2})
	-x_{2}(-j^{+}A_{1}+\pi^{-}\pa_{-}A_{1}+\pi^{i}\pa_{i}A_{1})
	-(\pi_{1}A_{2}-\pi_{2}A_{1})\}
\end{equation}
\begin{equation}
\nonumber
=\int dx^{-}d\bar{x}~\{ (x_{1}A_{2}-x_{2}A_{1})(\pa_{-}\pi^{-}+\pa_{i}\pi^{i}+j^{+})-(\pi_{2}A_{1}
	-\pi_{1}A_{2})-(\pi_{1}A_{2}-\pi_{2}A_{1})\}
\end{equation}
\begin{equation}
\label{dilc1}
=\int dx^{-}d\bar{x}~\{ (x_{1}A_{2}-x_{2}A_{1})
	(\pa_{-}\pi^{-}+\pa_{i}\pi^{i}+j^{+})\}.
\end{equation}
Partial integrations are done to obtain the last step which shows that the expressions are equal 
modulo Gauss constraint \eqref{lcg}. Rest of the components  of $	M_{+i}^{N}$ and $M_{+i}^{S}$  are given by,
\begin{equation}
	\label{lcan1}
	M_{+i}^{N}=\int dx^{-}d\bar{x} \left\{ (x_{+}T_{-i}-x_{i}T_{-+})-(\pi_{+}A_{i}-\pi_{i}A_{+})\right\},
\end{equation}
\begin{equation}
	\label{lcas1}
	M_{+i}^{S}=\int dx^{-}d\bar{x} \{ (x_{+}\Theta_{-i}-x_{i}\Theta_{-+})\}.
\end{equation}
Their  difference turns out to be,
\begin{equation}
\nonumber
	\label{lcamd1}
	M_{+i}^{N}-M_{+i}^{S}=\int dx^{-}d\bar{x} ~\{ x_{+}(T_{-i}-\Theta_{-i})-x_{i}(T_{-+}-\Theta_{-+})-(\pi_{+}A_{i}-
	\pi_{i}A_{+})\}
\end{equation}
\begin{equation}
\nonumber
=\int dx^{-}d\bar{x}~\{ (x_{+}A_{i}-x_{i}A_{+})(\pa_{-}\pi^{-}+\pa_{i}\pi^{i}+j^{+})-
	(\pi_{i}A_{+}
	-\pi_{+}A_{i})-(\pi_{+}A_{i}-
	\pi_{i}A_{+})\}\\
\end{equation}

\begin{equation}
\nonumber
=\int dx^{-}d\bar{x}~\{ (x_{+}A_{i}-x_{i}A_{+})
	(\pa_{-}\pi^{-}+\pa_{i}\pi^{i}+j^{+})\}.
\end{equation}
Likewise, the difference among the boosts defined as,
\begin{equation}
\label{lcbn}
M_{-i}^{N}=\int dx^-~d\bar{x}~ (x_{-}T_{-i}-x_{i}T_{--}-\pi_{i}A_{-})
\end{equation}
and,
\begin{equation}
\label{lcbs}
M_{-i}^{S}=\int dx^-~d\bar{x}~ (x_{-}\Theta_{-i}-x_{i}\Theta_{--})
\end{equation}
also turns out to be proportional to the Gauss constraint,
\begin{equation}
\label{difyo}
M_{-i}^{N}-M_{-i}^{S}= \int dx^{-} d\bar{x}(x_{-}A_{i}-x_{i}A_{-})(\pa_{-}\pi^{-}+\pa_{i}\pi^{i}+j^{+})
\end{equation}
Hence we have explicitly demonstrated that, in lightcone coordinates as well, the spacetime symmetry generators, obtained from the Noether and symmetric prescriptions, are equal modulo Gauss constraint which means that they are identical when acting on  the physical subspace. It is also straightforward to establish the energy momentum conservation for the fluid gauge model in lightcone framework where the fundamental brackets provided in (\ref{f},\ref{lcd}) need to be used. We have not given the detailed derivation since it is not very illuminating.

\section{Conclusion and future prospects}
As emphasized in the Introduction, our aim was to study in detail a first order Lagrangian field theory that is relativistic in nature and essentially depends on auxiliary field variables. These theories are primarily constrained systems and are structurally very distinct from conventional field theories that are generically quadratic and do not exploit auxiliary degrees of freedom. As an interesting and topical example we have chosen  the relativistic fluid model.

Let us now summarize our work. We have extended our previous work \cite{arpan} in two ways:    we have added the entropy term to the fluid sector and have included the Maxwell term in the gauge sector. The latter makes the gauge fields dynamical so that a fully interacting gauge-fluid theory  has been considered. We have concentrated primarily on the relativistic aspect of the theory and have studied in detail the structures of energy momentum tensor, derived from two definitions, {\it{ie.}} the canonical (Noether) one and the symmetric one. In the conventional equal-time formalism, we have shown that all the spacetime symmetry generators obtained from these two definitions agree modulo the Gauss constraint. This equivalence in the physical sector has been achieved only because of the kinetic term of the gauge fields. We consider this an important finding since, in the absence of this term, this equivalence cannot be shown. Subsequently we have explicitly demonstrated the validity of the Schwinger condition in the full theory. Apart from  it's intrinsic appeal it also ensures  that the unconventional nature of the fluid symplectic structure (with the  auxiliary fluid variables) does not spoil the relativistic covariance of the model.

Another important aspect of our work is the detailed analysis of the gauge-fluid model in the lightcone formalism. Inded this lightcone analysis has several non-trivial features but unfortunately a rigorous analysis of it, especially in the context of fluid dynamics has been lacking. Because of the recent interest in lightcone framework, we have carried out a detailed study. We have shown that the conservation principles are maintained. Furthermore we have explicitly demonstrated that as in the equal time case discussed here, the space time symmetry generators differ by the lightcone form of Gauss law.

What are the possible future directions? One interesting and topical problem is in the context of  anomalous fluid dynamics with triangle
anomalies in the form of Adler-Bell-Jackiw   anomaly \cite{ano, bel}. Of particular interest is the recent     work \cite{nair} where the authors propose a field theoretic fluid gauge model to represent  the   relativistic hydrodynamic formulation   of Abelian
gauge anomaly  in \cite{1son}. However, the work in \cite{nair} deals with a purely non-dynamical gauge field and we believe that it can be improved by introducing the Maxwell term in the gauge sector thereby studying the fully interacting theory as we have done here. Also the Hamiltonian analysis for the anomalous theory will be interesting both in equaltime and lightcone frameworks due to the presence of the non-trivial algebra in the gauge sector.

\end{document}